\documentclass[12pt,onecolumn,twoside]{IEEEtran}

\IEEEoverridecommandlockouts
\usepackage{amsmath ,amssymb ,euscript ,yfonts
,psfrag,latexsym,dsfont,graphicx,bbm,color,amstext,wasysym,pdfsync,balance,flushend}
\usepackage[usenames,dvipsnames]{xcolor}
\graphicspath{{./},{../figures/}{./figures/}}

\newtheorem{thm}{Theorem}

\newtheorem{prop}[thm]{Proposition}

\newtheorem{remark}[thm]{Remark}

\newcommand{\cS}{{\mathcal S}}

\newcommand{\cF}{{\mathcal F}}
\newcommand{\cM}{{\mathcal M}}
\newcommand{\cP}{{\mathcal P}}

\newcommand{\cA}{{\mathcal A}}
\newcommand{\cC}{{\mathcal C}}

\newcommand{\mC}{{\mathbb C}}

\newcommand{\mH}{{\mathbb H}}
\newcommand{\mN}{{\mathbb N}}
\newcommand{\mR}{{\mathbb R}}

\newcommand{\mI}{{\mathbb I}}
\newcommand{\bmu}{{\boldsymbol \mu}}
\newcommand{\bnu}{{\boldsymbol \nu}}
\newcommand{\mf}{{\boldsymbol f}}
\newcommand{\mw}{{\boldsymbol w}}
\newcommand{\bm}{{\boldsymbol m}}
\newcommand{\brho}{{\boldsymbol \rho}}
\newcommand{\boldf}{{\boldsymbol f}}
\newcommand{\blambda}{{\boldsymbol \lambda}}

\newcommand{\bphi}{{\boldsymbol \phi}}

\newcommand{\trace}{{\operatorname{tr}}}
\newcommand{\dd}{{\operatorname{d}}}

\def\spacingset#1{\def\baselinestretch{#1}\small\normalsize}

\title{\LARGE \bf Metrics for matrix-valued measures\\ via test functions}

\author{Lipeng Ning\thanks{L. Ning is with 
Brigham and Women's Hospital,
Harvard Medical School, Boston, MA 02115, 
{lning@bwh.harvard.edu}} and Tryphon T. Georgiou\thanks{T.T. Georgiou is with the Department of Electrical \& Computer Engineering,
University of Minnesota, Minneapolis, MN 55455, {tryphon@umn.edu}}\thanks{The research was supported in part by the NSF under Grant 1027696, the AFOSR under Grant FA9550-12-1-0319, and the Vincentine Hermes-Luh Endowment.}}

\begin{document}
\maketitle

\begin{abstract}
It is perhaps not widely recognized that certain common notions of distance between probability measures have an alternative dual interpretation which compares corresponding functionals against suitable families of test functions. This dual viewpoint extends in a straightforward manner to suggest metrics between matrix-valued measures. Our main interest has been in developing weakly-continuous metrics that are suitable for comparing matrix-valued power spectral density functions. To this end, and following the suggested recipe of utilizing suitable families of test functions, we develop a weakly-continuous metric that is analogous to the Wasserstein metric and applies to matrix-valued densities. We use a numerical example to compare this metric to certain standard alternatives including a different version of a matricial Wasserstein metric developed in \cite{Ning2013matrix, Ning2013}.
\end{abstract}

\spacingset{1.07}
\section{Introduction}
Consider the set of probability measures 
\[
\cP(\mI):=\left\{\mu : d\mu(\theta)\geq0 \text{~for~} \theta \in \mI \text{~and~} \mu(\mI)=1 \right\}
\]
where, herein, $\mI$ is always thought to be an interval. 
There is a large family of metrics, comparing $\mu_1, \mu_2\in \cP$, that are expressed in the form
\begin{align}\label{eq:dist}
\sup_{f\in\cF} |\int f(d\mu_1-d\mu_2)|
\end{align}
with $\cF$ being a suitable set of functions on $\mI$.
Probability metrics that can be expressed in this form \eqref{eq:dist} are often referred to either as metrics with a $\zeta$-structure \cite{Zolotarev1984probability} or as integral probability metrics \cite{Muller1997integral,rachev2013methods}.

The family of metrics that can be expressed as in \eqref{eq:dist} includes many familiar ones such as the total variance, Kolmogorov's distance and the 1-Wasserstein metric. 
In point of fact, the total variation which is defined by
\[
\|\mu_1-\mu_2\|_{\rm TV}:=\int_{\mI}|d\mu_1(x)-d\mu_2(x)|,
\]
can also be expressed as
\[
\|\mu_1-\mu_2 \|_{\rm TV}=\sup_{f}  \left\{ \int f(d\mu_1-d\mu_2) \mid \|f\|_{\infty}\leq 1\right\},
\]
Kolmogorov's metric which is defined by
\[
\dd_{\rm K}(\mu_1, \mu_2):=\sup_{x\in \mI} |F_1(x)-F_2(x)|
\]
with $F_1$ and $F_2$ being the cumulative distribution function of $\mu_1$ and $\mu_2$, respectively, can be expressed as 
\[
\dd_{\rm K}(\mu_1, \mu_2)=\sup_{f} \left\{ \int_{\mI} f (d\mu_1-d\mu_2) \mid   \int_{\mI}|f^\prime|dx\leq 1\right\}
\]
see \cite[page 73]{Rachev1991probability}, and so does the $1$-Wasserstein metric \cite{Villani_book}.

We focus on the $1$-Wasserstein metric which can be defined over more general spaces. It is defined with respect to the metric $d(x,y)=|x-y|$, as follows. If
\[
\cP_{\mu_1, \mu_2}\subset \cP(\mI\times \mI)
\]
denotes the subset of probability measures on $\mI\times \mI$ that have $\mu_1$ and $\mu_2$ as marginals, the $1$-Wasserstein distance between $\mu_1$ and $\mu_2$ is 
\begin{equation}\label{eq:primal}\nonumber
\dd_{W_1}(\mu_1, \mu_2):=\inf_{m\in \cP_{\mu_1, \mu_2}}  \int_{\mI\times \mI} d(x,y) dm(x, y) .
\end{equation}
Naturally, it can be expressed in a dual form as
\begin{equation}\label{eq:dual}
\dd_{W_1}(\mu_1, \mu_2)=\sup_{f} \left\{ \int_{\mI} f (d\mu_1-d\mu_2) \mid   \|f\|_{\rm Lip}\leq 1\right\}
\end{equation}
where 
\[
\|f\|_{\rm Lip}:=\sup_{x,y\in \mI} \frac{|f(x)-f(y)|}{|x-y|}
\]
denotes the Lipschitz semi-norm.

An important property of the Wasserstein metric is that it metrizes weak$^*$ convergence of probability measures
\cite[page 212]{Villani_book}, that is, given $\mu$ and
any sequence of measures $\{\mu_k, k=1, 2, \ldots \}$,
\[
\dd_{W_1}(\mu_k, \mu)\underset{k\rightarrow \infty}{\longrightarrow} 0
\]
if and only if, for all continuous and bounded $f$,
\[
\int_{\mI} fd\mu_k \underset{k\rightarrow \infty}{\longrightarrow} \int_{\mI} fd\mu.
\] 
Intuitively, small changes in a weak$^*$-sense reflect small changes in any relevant statistics. This is clearly a desirable feature in any experimental engineering quantification of distances between distributions, whether these represent probability, power, spectral power or other entities.
For this precise reason, the Wasserstein metric has turned out to be a useful tool in modeling of slowly varying time-series \cite{Jiang2012Geometric} and for comparing covariance matrices \cite{Ning2013Covariance}, among many other applications \cite{Rachev1991probability}.

In the sequel we say that a metric is {\em weakly continuous} if it metrizes weak$^*$ convergence. The use of the terminology  ``weak'' instead of ``weak$^*$'' is common in probability theory and is adapted in this paper.

The main contribution of the present paper is a particular generalization of the Wasserstein metric to the space of matrix-valued measures with possibly non-equal mass. This metric differs from an analogous metric in our recent work \cite{Ning2013matrix} which also represents a generalization of Wasserstein distances to matricial measures.
The present metric is based on a dual formalism where we compare measures on a suitable set of test functions.
Before dealing with the matricial case, in Section \ref{sec:scalar}, we first discuss how to modify the Wasserstein metric so as to compare measures with non-equal masses; a subsequent matricial generalization follows along similar lines. In Section \ref{sec:non-commutative}, we present certain related ideas from non-commutative geometry for devising metrics to compare states of non-commutative algebras; such states are ``non-commutative'' generalizations of probability measures. In Section \ref{sec:matrix}, we develop the sought weakly-continuous metric between matrix-valued measures. Our interest is in spectral analysis of multivariate time-series and, thereby, we appeal to a pertinent numerical example to highlight differences and similarities of the proposed metric to alternatives; this is given in Section \ref{sec:Example}.

The notational convention we follow is to use regular font, as in $\mu, m$, for scalar values, variables, and functions, and to use boldface fonts, as in $\bmu, \bm$, for matrix-valued functions or elements of a general algebra.

\section{Wasserstein-like metric for unbalanced measures}\label{sec:scalar}

We begin by discussing a certain adaptation of the Wasserstein metric to use on unbalanced measures  \cite{Georgiou2009metrics}, that is, for the case when the measures we deal with may have unequal integrals.
Our motivation stems from the need for a weakly continuous distance to be used on power spectral densities of stationary stochastic processes. The dual formulation serves as a template for a subsequent matricial version.

Consider the set of non-negative scalar measures 
\[
\cM(\mI):=\left\{\mu : d\mu(\theta)\geq0 \text{~for~} \theta \in \mI\right\}.
\]
For any two $\mu_1, \mu_2\in\cM$, let
\begin{align}\label{eq:dwk}
&\dd_{{\rm W}_1, \kappa}(\mu_1, \mu_2):=\nonumber\\
&\inf_{\hat\mu_1,\hat \mu_2}\left\{ \dd_{W_1}(\hat\mu_1,\hat\mu_2)+\kappa \sum_{k=1}^2 \|\mu_k-\hat\mu_k\|_{\rm TV}\right\}
\end{align}
where $\kappa>0$ is used to weigh in the relative importance of the two terms. In \cite{Georgiou2009metrics}, the optimizing variables $\hat\mu_1, \hat\mu_2$ represent ``noise-free'' measures having equal mass while the ``error'' differences $\mu_k-\hat\mu_k$ for $k=1, 2$, are attributed to statistical variability. The dual of \eqref{eq:dwk} is
\begin{align}\label{eq:dwk_dual}
&\dd_{{\rm W}_1, \kappa}(\mu_1, \mu_2)=\nonumber\\
&\hspace*{-.3cm}\sup_{f}\{ \int_{\mI} f(d\mu_1-d\mu_2) \mid \|f\|_{\rm Lip}\leq 1, \|f\|_{\infty}\leq \kappa \}
\end{align}
where $\|f\|_{\infty}=\max_{x\in \mI} |f(x)|$.
Since test functions in \eqref{eq:dwk_dual} are bounded, $\dd_{{\rm W}_1, \kappa}(\mu_1, \mu_2)$ is bounded as well
\cite{Georgiou2009metrics}, thereby, it can be easily shown that $\dd_{{\rm W}_1, \kappa}$ is a weakly continuous metric (see \cite{Georgiou2009metrics} for details). Applications of this metric to power spectral analysis has been pursued in \cite{Karlsson2012uncertainty}.

We remark that the essence in \eqref{eq:dist} is to postulate a family of test functions which is rich enough so as to distinguish measures while the functions in the class be equicontinuous and uniformly bounded \cite{Karlsson2012uncertainty}. For the Wasserstein metric, the family of test function is the class of Lipschitz functions. 
A very similar rationale has been introduced in non-commutative geometry, where via a suitable generalization of the Lipschitz semi-norm, one obtains a metric between non-commutative states, namely the Connes' spectral distance. This is discussed next.

\section{Metrics in non-commutative geometry}\label{sec:non-commutative}

We specialize our discussion to the algebra of $n\times n$ complex-valued matrices $M_n(\mC)$; for the more general setting of non-commutative algebras see \cite{Connes1995noncommutative,Rieffel1999metrics}.
Given $A, B\in M_n(\mC)$, or any algebra, we denote by $[A,B]=AB-BA$ the commutator of $A$ and $B$.

In general, a state $\brho$ of a non-commutative algebra $\cA$ is a positive linear functional that maps $f\in \cA$ to $\mR$ or $\mC$. The set of states of $\cA$ is denoted as $\cS(\cA)$. For example, a state $\brho$ of the commutative algebra $\cC(\mI)$ of continuous functions on $\mI$ uniquely corresponds to a probability measure $\mu$ such that for any $f\in \cC(\mI)$
\[
\brho(f)=\int_{\mI} f d\mu. 
\]
In quantum mechanics, a state $\brho$ of a non-commutative algebra of ``observables'' is a {\em density matrix}; this is positive semidefinite with trace equals to one. For $\boldf\in \cA$,
\[
\brho(\boldf)=\trace(\brho \boldf).
\]
Thus, states are thought of as a generalization of probability measures and $\brho(\cdot)=E_\brho\{\cdot\}$ is thought of as the expectation operator.

\subsection{Connes' spectral distance}

Consider a non-commutative algebra of operators $\cA$ on a Hilbert space. Let $D$ be a specific fixed operator often referred to as the Dirac operator. Connes' spectral distance \cite{Connes1995noncommutative} between $\brho_1, \brho_2\in \cS(\cA)$ is defined as
\begin{align*}
\dd_{D}(\brho_1, \brho_2):=\sup_{\boldf\in\cA} \left\{|\brho_1(\boldf)-\brho_2(\boldf) |  \mid  \|[D, \boldf]\| \leq 1\right\}.
\end{align*}

The operator norm of the commutator $[D, \boldf]$ takes the role of the Lipschitz semi-norm. 
To illustrate the insight in viewing $\|[D, \boldf]\|$ as the analogue of a Lipschitz semi-norm, consider $f$ to be a smooth function on $\mR$ and choose $D=-i\partial_x$. For any smooth function $g$,
\begin{align*}
[D, f]g&=Dfg-fDg\\
&=-i\partial_x(fg)+if\partial_xg\\
&=-i\partial_x f g.
\end{align*}
Thus $[D, f]$ takes $g\mapsto -i\partial_x fg$. The operator norm $\|[D, f]\|$ equals to the operator norm the mapping $g\mapsto -i\partial_x fg$, which is precisely the Lipschitz semi-norm of $f$.

Consider $\cA$ to be the algebra $\cC(\mI)$ of continuous functions on $\mI$. Thus, for any two $\brho_1, \brho_2\in \cS(\cC(\mI))$ which correspond to probability measures $\mu_1$ and $\mu_2$, respectively,
\[
\dd_D(\brho_0, \brho_1)=\sup_{\boldf\in\cA} \left\{ \int_{\mI} f(d\mu_1- d\mu_2)  \mid  \|[D, f]\| \leq 1\right\}.
\]
The relation of this distance to the  1-Wasserstein is explained in \cite{Rieffel1999metrics,DAndrea2010view}.
We elaborate with an example and provide directions for further generalization.

Connes' spectral distance can be unbounded. For instance, let $\cA=M_2(\mR)$ and let 
\[
D=\left[\begin{matrix}0&1\\1& 0\end{matrix} \right].
\]
Consider as states the set  $\cS(\cA)$ of positive semi-definite matrices with trace equal to one ---observables can also be taken in $\cA$. Take an observable
\[
\boldf=\left[\begin{matrix}a&b\\c& d\end{matrix} \right]
\]
and note that
\[
[D, \boldf]=\left[\begin{matrix}c-b&d-a\\ a-d & b-c\end{matrix} \right].
\] 
Thus, the distance between states
\[
\brho_0=\left[\begin{matrix}p_0&q_0\\q_0& 1-p_0\end{matrix} \right] \text{~and~}
\brho_1=\left[\begin{matrix}p_1&q_1\\q_1& 1-p_1\end{matrix} \right],
\]
namely,
\begin{align*}
&d_{D}(\brho_0, \brho_1)\\
&=\sup_{\boldf\in \cA}\left\{|\trace(\brho_0\boldf)-\trace(\brho_1\boldf)| ~\mid~ \|[D, \boldf]\|\leq 1 \right\}\\
&=\sup\bigg\{|(p_0-p_1)(a-d)+(q_0-q_1)(b+c)| \\
&\hspace*{1.5cm}~\mid~ \|\left[\begin{matrix}c-b&d-a\\ a-d & b-c\end{matrix} \right]\|\leq 1 \bigg\}
\end{align*}
is $\infty$. To see this note that
if $q_0\neq q_1$, then $d_{D}(\brho_0, \brho_1)$ can take any positive value for a suitable choice of $b=c$.

\begin{remark}
In analogy with \eqref{eq:dwk_dual}, 
\begin{align}\label{eq:dwk_dual2}
&\dd_{D,\kappa}(\brho_1, \brho_2):=\\
&\sup_{\boldf\in\cA} \bigg\{|\brho_1(\boldf)-\brho_2(\boldf) | \mid  \|[D, \boldf]\| \leq 1, \| \boldf\|\leq \kappa\bigg\}\nonumber
\end{align}
defines a bounded metric. Likewise, the Connes' spectral distance can be readily generalized to
\begin{align}\nonumber
\dd_D(\brho_0, \brho_1)&=\sup_{\boldf\in\cA} \left\{  \phantom{\big|}|\brho_1(\boldf)-\brho_2(\boldf) | \mid  \|[D_i, \boldf]\| \leq 1\right.\\
&\left.\phantom{\big| xxxxxx}\mbox{ for } i\in\{1,2\,\ldots, n\}\right\}\label{eq:dwk_dual3}
\end{align}
for a suitable set of $n$ Dirac operators quantifying ``slope''  in several possible ``directions.''
\end{remark}

In the next section we deal in more detail with the algebra of matrix-valued continuous functions on $\mI$, namely, $\cC(\mC^{n\times n},\mI)$ where states correspond to matrix-valued measures. Our interest stems from spectral analysis of multivariate time-series and in the next section, we present a Wasserstein-like metric between matrix-valued measures. The formalism is completely analogous in that the metric is constructed in a dual formalism by quantifying how measures act on suitably constrained matrix-functions.

\section{A Wasserstein-like metric between matrix-valued measures}\label{sec:matrix}
Spectral analysis of time-series aims at detecting power and correlations between signals at different parts of the frequency spectrum. Power spectral estimates are typically based on moments, i.e., integrals of the power density, or simply measurements. Hence, it is unreasonable to utilize metrics between power spectral densities that are not weakly continuous.

For the case of multivariable time-series, power densities are matrix-valued. Thus, any metric must weigh in both the frequency content of the power as well as its directionality. Typically, the directionality of singular vectors of a matrix-power spectral density at a given frequency relates to the relative strength of the corresponding signal-components at the location of the measurement channels (sensors). Therefore, in order to quantify the performance of estimation algorithms, accurately detect changes in time series (events), localize the directionality of echo (e.g., in radar), etc., one needs physically meaningful metrics that weigh in relevant characteristics of power spectra. Thus, at the very least, metrics ought to be weakly continuous and allow the comparison of matrix-valued densities.

Several distance measures have been proposed and extensively used in applications and the literature, see e.g., \cite{Ferrante_time,Ferrante_hellinger,Ferrante2012,Afsari2013,JNG,Ning2013matrix}. However, besides the fact that these fail to be metrics, most fail to be weakly continuous as well (e.g., the Itakura-Saito distance, etc.). In the present section, following the recipe outlined earlier, in (\ref{eq:dwk_dual}-\ref{eq:dwk_dual3}), we develop a weakly continuous metric between matrix-valued measures that can be thought of as a generalization of the Wasserstein metric. Our viewpoint herein differs from, yet it can be seen as complementing to the one in \cite{Ning2013matrix}.

For a Hermitian matrix-valued function $\boldf$, let
\[
\|\boldf\|_{\rm Lip}:=\sup_{x\neq y} \frac{\|\boldf(x)-\boldf(y) \|}{d(x,y)}
\]
where $d(x,y)$ is a metric on $\mI$. Then $\|\boldf\|_{\rm Lip}\leq 1$ implies that \[
\boldf(x)-\boldf(y)\leq d(x,y)I
\]
with $I$ being the identity matrix
in the sense of positive semi-definiteness. For any two matrix-valued power spectral measures $\bmu_1, \bmu_2$ we define:
\begin{align}\label{eq:dwk_dual_matrix}
&\dd_{{\rm W}_1, \kappa}(\bmu_1, \bmu_2)=\nonumber\\
&\sup_{\boldf}\{ \int_{\mI} \trace(\boldf(d\bmu_1-d\bmu_2) \mid \|\boldf\|_{\rm Lip}\leq 1, \|\boldf\|\leq \kappa \}.
\end{align}
It is quite clear that this defines a metric and as we state next, it is also weakly continuous.

\begin{prop}\label{prop:MatrixWeakContinuity}
Consider a sequence of matrix-valued power spectral measures $\{\bmu_k : k=1, 2, \ldots\}$ and $\bmu$. Then
\begin{equation}\label{eq:MetricConv}
\dd_{{\rm W}_1,\kappa}(\bmu_k, \bmu)\underset{k\rightarrow \infty}{\longrightarrow} 0
\end{equation}
if and only if for any continuous, bounded, Hermitian-valued function $\boldf$ on $\mI$ the following holds
\begin{equation}\label{eq:SequenceConv}
 \trace\left( \int_{\mI}\boldf d\bmu_k\right)\underset{k\rightarrow \infty}{\longrightarrow}\trace\left( \int_{\mI} \boldf d\bmu\right).
\end{equation}
\end{prop}
\vspace*{.1in}

\begin{proof}
See Appendix \ref{Appendix1}.
\end{proof}
\vspace*{.1in}

We provide an interpretation of \eqref{eq:dwk_dual_matrix} that draws a connection to optimal mass transport very much like in Section \ref{sec:scalar}. For this, we need the following expression for the total variation:
\[
\|\bmu_1-\bmu_2\|_{\rm TV}:=\int_{\mI} \|d\bmu_1-d\bmu_2\|_*
\]
where $\|\cdot\|_*$ denotes the nuclear norm, i.e., the sum of singular values.
We also need to the following matricial analog of the 1-Wasserstein metric between matrix-valued measures $\bmu_1$ and $\bmu_2$ with the same ``total matricial mass'' $\bmu_1(\mI)=\bmu_2(\mI)$,
\begin{align*}
&\dd_{W_1}(\bmu_1, \bmu_2)=\inf_{\bm} \bigg\{\int_{\mI\times \mI} d(x,y) \|d\bm(x,y)\|_*\mid\nonumber\\
&\int_{y\in \mI}d\bm(x,y)=d\bmu_1(x), \int_{x\in \mI}d\bm(x,y)=d\bmu_2(y) \bigg\}.\nonumber
\end{align*}
We remark that, in the above, $d\bm(x,y)$ needs not to be positive semidefinite. The optimization seeks a distribution for the nuclear norm of $d\bm(x,y)$ which may now be thought of actually as the ``mass" transported from $x$ to $y$. 
When $\bmu$ and $\bnu$ are scalar-valued probability measure, clearly, $\dd_{W_1}(\bmu,\bnu)$ is the 1-Wasserstein metric. The following statement represents a generalization of \eqref{eq:dwk}.

\begin{prop}\label{prop:PrimalForm}
For two matrix-valued measures $\bmu_1$ and $\bmu_2$ on $\mI$, 
\begin{align}\label{eq:dwk_matrix}
&\dd_{{\rm W}_1, \kappa}(\bmu_1, \bmu_2):=\\
&\inf_{\hat\bmu_1,\hat \bmu_2} \left\{\dd_{W_1}(\hat\bmu_1,\hat\bmu_2)+\kappa \sum_{k=1}^2 \|\bmu_k-\hat\bmu_k\|_{\rm TV} \right\}\nonumber
\end{align}
\end{prop}
\vspace*{.1in}

\begin{proof}
See Appendix \ref{Appendix2}.
\end{proof}

\section{Example}\label{sec:Example}
We highlight the characteristics of the proposed distance $\dd_{{\rm W}_1, \kappa}$ with a numerical example. In this, we compare three matrix-valued densities $\mf_0, \mf_1, \mf_2$ and compute the values assigned by the metric between them, for each pair. The nature and directionality of their spectral content is such that $\mf_1$ can naturally be thought of as ``sitting'' in the middle of the other two. Thus, if a metric is to be intuitive, it ought to assign distances accordingly. We compare the relative values assigned by our metric as well as three alternatives. These are, $d_{\rm IS}$, $d_{\rm TV}$ and the metric introduced in our earlier work \cite{Ning2013matrix}. It is seen that \eqref{eq:dwk_dual_matrix}, as well as the metric in \cite{Ning2013matrix}, are quite similar, while the other two, 
 $d_{\rm IS}$ and $d_{\rm TV}$, give relative distances that do not reflect the intuition suggested above.\\[-.09in]

\begin{figure}[htb]\begin{center}
\includegraphics[totalheight=9cm]{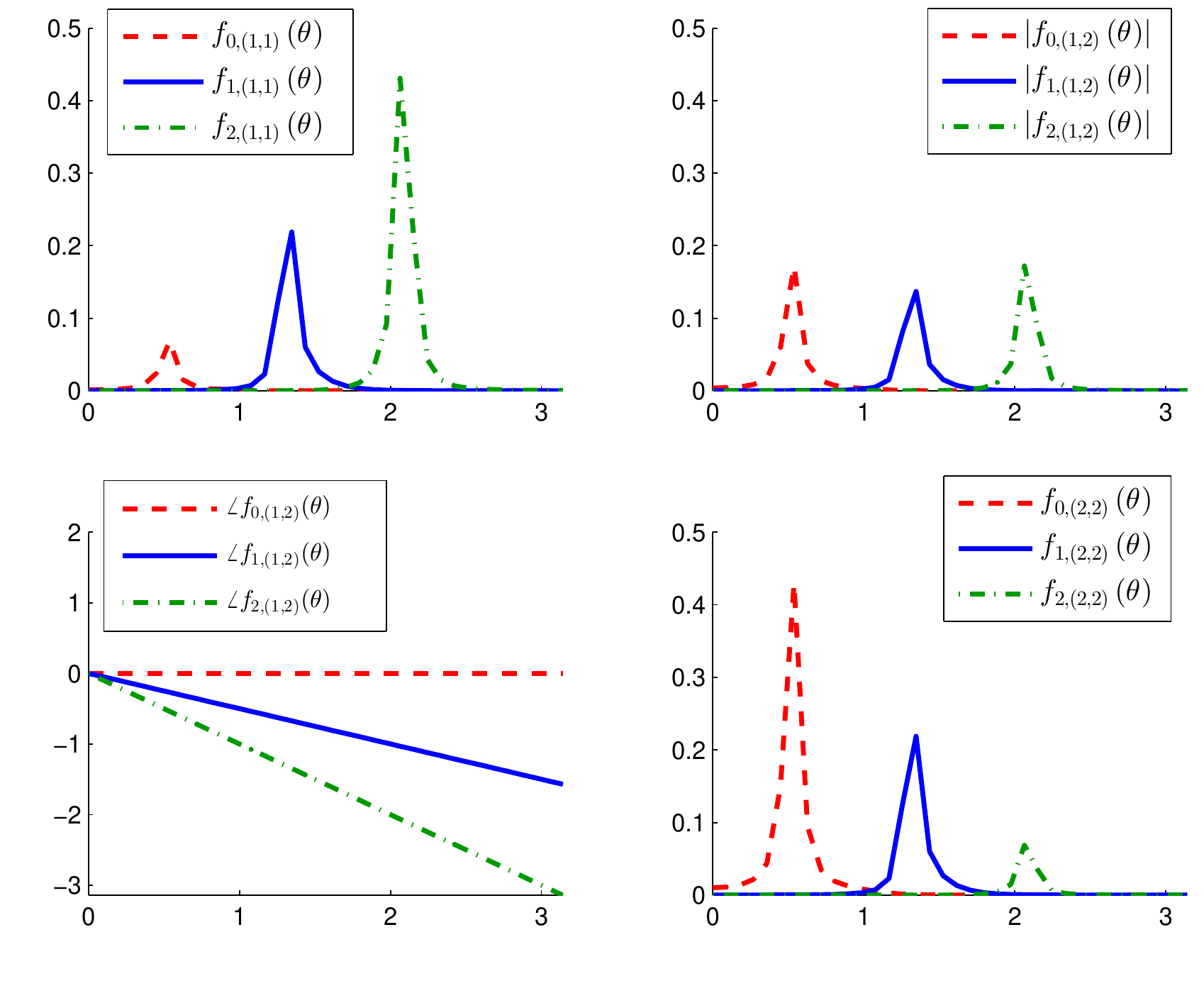}
\caption{Matrix-valued power spectra $\mf_0$, $\mf_1$ and $\mf_2$ are shown in red dashed line, blue solid line and green dashdot line, respectively.
Subplots (1,1), (1,2) and (2,2) show $\mf_{i,(1,1)}, |\mf_{i,(1,2)}|$ (same as $|\mf_{i,(2,1)}|$) and $ \mf_{i,(2,2)}$. Subplot (2,1) shows $\angle(\mf_{i,(1,2)})$ for $i\in\left\{ 0,1, 2\right\}$.}\label{fig:MatrixSpectra}\end{center}
\end{figure}

The three chosen matricial densities  are:
{\footnotesize
\begin{align*}
\mf_0(\theta)&=\left[\begin{matrix} ~1~&~0.4~\\~0& ~1~\end{matrix} \right]\left[\begin{matrix}~~0.01~~&~0~\\~0~& \frac{1}{|a_0(e^{j \theta})|^2}\end{matrix} \right]\left[\begin{matrix} ~1~&~0~\\~0.4~& ~1~\end{matrix} \right]\\\ \\
\mf_1(\theta)&=\left[\begin{matrix}1&0.5\\0.5e^{j \theta}& 1\end{matrix} \right]\left[\begin{matrix}\frac{1}{|a_1(e^{j \theta})|^2}&0\\0& \frac{1}{|a_1(e^{j \theta})|^2}\end{matrix} \right]\left[\begin{matrix}1&0.5e^{-j \theta}\\0.5& 1\end{matrix} \right]\\\\
\mf_2(\theta)&=\left[\begin{matrix}~1~&~0~\\0.4e^{j \theta}& ~1~\end{matrix}\right]\left[\begin{matrix}\frac{1}{|a_2(e^{j \theta})|^2}&~0~\\~0~& ~0.01~\end{matrix} \right]\left[\begin{matrix}\;~1~\;&0.4e^{-j\theta}\\~0~& ~1~\end{matrix} \right]
\end{align*}
}where
{\footnotesize
\begin{align*}
a_0(z)&=(1-1.9\cos(\frac{\pi}{6})z+0.95^2 z^2 )(1-1.5\cos(\frac{\pi}{3})z+0.75^2 z^2 )\\
a_1(z)&=(1-1.9\cos(\frac{5\pi}{12})z+0.95^2 z^2 )(1-1.5\cos(\frac{\pi}{2})z+0.75^2 z^2 )\\
a_2(z)&=(1-1.9\cos(\frac{2\pi}{3})z+0.95^2 z^2 )(1-1.5\cos(\frac{5\pi}{8})z+0.75^2 z^2 )
\end{align*}
}for $\theta\in[0, \pi]$.
These are shown in Figure \ref{fig:MatrixSpectra}
and since, for each $i\in\{0,1,2\}$, $\mf_i$ is a Hermitian-valued and positive semi-definite, our convention is to display the magnitude and phase of their entries as follows: $|\mf_{i,(1,1)}|$, $|\mf_{i,(1,2)}|$ ($=|\mf_{i,(2,1)}|$) and $|\mf_{i,(2,2)}|$ are displayed in subplots (1,1), (1,2) and (2,2), respectively,
and $\angle \mf_{i,(1,2)}$  ($= - \angle \mf_{i,(2,1)}$) is shown in subplot (2,1).

Peak power in $\mf_0$ is at $\theta=\frac{\pi}{6}$ and most of it resides in the second channel, i.e., in $\mf_{0,(2,2)}$.
The power in $\mf_1$ splits equality between the two channels and its peak is at $\theta=\frac{5\pi}{12}$. In $\mf_2$, peak power is in the first channel and at around $\theta=\frac{2\pi}{3}$. Similarly, it is worth observing the characteristics and phase angles of the cross spectra.
All in all, $\mf_1$ appears to be ``sitting in the middle'' between $\mf_0$ and $\mf_2$. By comparing the relative distances we can now assess whether these reflect the above intuition, i.e., that $\mf_1$ is in the middle between $\mf_0$ and $\mf_2$.

We compute and compare distances given by the generalized Itakura-Saito distance \cite{Ferrante_time}
\begin{equation*}
\dd_{\rm IS}(\mf_0,\mf_1):=\int_0^{\pi} \trace\left( \mf_0\mf_1^{-1}-\log (\mf_0\mf_1^{-1})-I \right)d\theta
\end{equation*}
the total variation $\dd_{\rm TV}(\mf_0, \mf_1):=\|\mf_0-\mf_1\|_{\rm TV}$ and the metric $\dd_{{\rm T},\kappa}$ proposed in \cite{Ning2013matrix}. To this end, we sample the $\mf_i$'s on a frequency grid with resolution $\Delta \theta=\frac{\pi}{35}$. To obtain $\dd_{{\rm T},\kappa}$, we first scale so that traces are normalized to $1$.
The distances are given in Table \ref{tab_distance} where we have chosen $\kappa=1$ for both $\dd_{{\rm W}_1, \kappa}$ and $\dd_{{\rm T}, \kappa}$.

\begin{table}[htb]
\centering
\begin{tabular}{|c|p{2cm}|p{2cm}|p{2cm}|}
  \hline
  & \multicolumn{3}{|c|}{Distance between pairs of density functions}  \\
  \cline{2-4}
  & $\mf_0, \mf_1$ & $\mf_1, \mf_2$ & $\mf_0, \mf_2$\\
   \hline
  $\dd_{\rm IS}$&$3.44\times 10^3$&$5.36\times 10^4$&$9.27\times 10^4$ \\ \hline
  $\dd_{\rm TV}$&1.95&1.96&2.00\\ \hline
  $\dd_{{\rm W}_1, \kappa}$& 1.37 &1.65  &2.29 \\ \hline
  $\dd_{{\rm T}, \kappa}$& 1.01 &1.09  &2.05 \\ \hline
\end{tabular}
\caption{Distances between density functions.}
\label{tab_distance}
\end{table}

We observe that $\dd_{\rm IS}$ gives a rather unintuitive result since we expect similar distances between the two pairs $\mf_0, \mf_1$ and $\mf_1, \mf_2$.
The total variation does not differentiate the three pairs since all three are found equally close to each other. Using $\dd_{{\rm T}, 1}$, $\dd_{{\rm T}, 1}(\mf_0, \mf_1)$ is very close to $\dd_{{\rm T}, 1}(\mf_1, \mf_2)$ and they are both at nearly half compared to $\dd_{{\rm T}, 1}(\mf_0, \mf_2)$. Then $\dd_{{\rm W}_1, 1}$ is quite similar to $\dd_{{\rm T}, 1}$.
These comparisons suggest that $\dd_{{\rm W}_1, \kappa}$ as well as $\dd_{{\rm T}, 1}$, both reflect quite closely the intuition and what one would expect after inspecting the relative distribution of power and directionality in all three densities.

\section{Concluding remarks}
The main thesis of this paper is that it is natural to quantify distances by evaluating measures against suitable families of test functions. After explaining the general recipe on representative metrics on scalar densities, including the one proposed in our earlier work \cite{Georgiou2009metrics}, we expand on possible directions that lead to metrics for matrix-valued measures and density functions. We discuss in detail one so-derived weakly continuous metric which can be thought of as a natural generalization of the 1-Wasserstein metric between matrix-valued densities. Comparison with alternatives as well as a similar metric in  \cite{Ning2013matrix} is explained on an academic example. A key property of the metric introduced herein is the weak continuity which is not shared by any of the alternatives.

\appendices
\section{Proof of Proposition \ref{prop:MatrixWeakContinuity}}\label{Appendix1}

First, it is clearly \eqref{eq:SequenceConv} implies \eqref{eq:MetricConv}. We show that \eqref{eq:MetricConv} also implies \eqref{eq:SequenceConv}. The proof below follows that in \cite[page 216]{Villani_book} for the scalar-valued Wasserstein metric.

We notice that if \eqref{eq:MetricConv} holds, then \eqref{eq:SequenceConv} hold for any Lipschitz and bounded $\boldf$.
To see this, for any Lipschitz and bounded $\boldf$, it can be scaled so that $\boldf/\max\{\|\boldf\|_{\rm Lip}, \frac{\max\|\boldf\|}{\kappa} \}$ is a feasible element in \eqref{eq:dwk_dual_matrix}. Then, the proof requires showing that any element in the set of continuous, bounded Hermitian matrix-valued function $C_b(\mH^{\ell\times \ell}, \mI)$ can be approached by sequences of Lipschitz and bounded ones from above and below respectively.

We need the following expression for suitable $\inf$ and the $\sup$ of an $\boldf \in C_b(\mH^{\ell\times \ell}, \mI)$:
\begin{eqnarray*}
\inf_{x\in \mI} \boldf(x) &:=&\arg\underset{\bm}{\sup}\left\{ \trace(\bm) \mid \boldf(x)- \bm\geq0\right\}\\
\sup_{x\in \mI} \boldf(x) &:=&\arg\underset{\bm}{\inf}\left\{ \trace(\bm) \mid  \bm-\boldf(x)\geq 0\right\}
\end{eqnarray*}
with ordering in the sense of positive semi-definiteness.
For $n\in\mN$, we denote
\begin{eqnarray*}
\boldf_{{\rm low}, n}(x) &:=&\inf_{y\in \mI} \left\{\boldf(y)+nd(x,y)I\right\}\\
\boldf_{{\rm up}, n}(x) &:=&\sup_{y\in \mI} \left\{\boldf(y)-nd(x,y)I\right\}.
\end{eqnarray*}
Then, $\boldf_{{\rm low}, n}(x)\leq \boldf(x)\leq \boldf_{{\rm up}, n}(x)$.
For any fixed $x$, $\{\boldf_{{\rm low}, n}(x) : n\in\mN\}$ and $\{\boldf_{{\rm up}, n}(x) :  n\in \mN \}$ are increasing sequences and decreasing sequences, respectively, and they both converge to $\boldf(x)$.
It is also important to note that $\|\boldf_{{\rm low}, n}\|_{\rm Lip}\leq n$ and $\|\boldf_{{\rm up}, n}\|_{\rm Lip}\leq n$. To see this, for any $x, y \in[0, 1]$, we have
\begin{align*}
&\boldf_{{\rm low}, n}(x)-\boldf_{{\rm low}, n}(y)\\
&\hspace*{-.2cm}=\inf_{z} \big\{\boldf(z)+nd(x,z)I\big\}-\inf_{\hat z} \big\{\boldf(\hat z)+nd(y,\hat z)I\big\}\\
&\hspace*{-.2cm}\leq \sup_{z}\big\{(\boldf(z)+nd(x,z)I)- (\boldf(z)+nd(y,z)I)\big\}\\
&\hspace*{-.2cm}=\sup_{z} n(d(x,z)- d(y,z))I\\
&\hspace*{-.2cm}\leq nd(x,y)I.
\end{align*}
Thus $\|\boldf_{{\rm low}, n}\|_{\rm Lip}\leq n$, and $\|\boldf_{{\rm up}, n}\|_{\rm Lip}\leq n$ can be proved in a similar manner.
Then we have
\begin{align*}
&\limsup_{k\rightarrow \infty} \trace\left(\int_0^1 \boldf  d\bmu_k \right)\\
&\leq \liminf_{n\rightarrow \infty} \limsup_{k\rightarrow \infty}  \trace\left(\int_0^1 \boldf_{{\rm up},n} d\bmu_k\right)\\
&=\liminf_{n\rightarrow \infty} \trace\left(\int_0^1 \boldf_{{\rm up},n} d\bmu\right)\\
&=\trace\left(\int_0^1 \boldf d\bmu\right).
\end{align*}
Similarly, using the sequence $\{\boldf_{{\rm low}, n}(x) : n\in\mN\}$ we can also show that
\[
\liminf_{k\rightarrow \infty} \trace\left(\int_0^1 \boldf d\bmu_k\right)\geq \trace\left(\int_0^1 \boldf d\bmu\right).
\]
This completes the proof.

\section{Proof of Proposition \ref{prop:PrimalForm}}\label{Appendix2}
We show that \eqref{eq:dwk_dual_matrix} is dual of \eqref{eq:dwk_matrix}: 
rewrite \eqref{eq:dwk_matrix} as
\begin{align}\label{eq:DwkExpand}
&\inf_{\bm, \hat\bmu, \hat\bnu} \bigg\{\int_{\mI\times \mI}  \hspace*{-5pt}d(x,y)\|\bm(x,y)\|_*+\kappa\sum_{k=1}^2 \|\bmu_k-\hat\bmu_k\|_{\rm TV} ~\mid\nonumber\\
 &\hspace*{-5pt}\int_{y\in\mI} \hspace*{-5pt}d\bm(x,y)=d\hat \bmu_1(x),~\int_{x\in \mI}  \hspace*{-5pt}d\bm(x,y)=d\hat \bmu_2(y) \bigg\}.
\end{align}
To derive the dual formulation, we need to rewrite the nuclear norm as follows
\begin{align*}
\|d\bm\|_*&=\max_{\|\mw\|\leq 1}\trace(\mw d\bm)\\
\|d\bmu_k-\hat d\bmu_k\|_*&=\max_{\|\blambda_k\|\leq 1}\trace(\blambda_k(d \bmu_k-d\bmu_k)).
\end{align*}
Let $\bphi_1, \bphi_2$ be the Lagrange multipliers for the two constraints. Using the Lagrange multiplier method, \eqref{eq:DwkExpand} equals the following
\begin{subequations}
\begin{align}
&\hspace*{-.2cm}\inf_{\bm, \hat \bmu_1, \hat \bmu_2}\hspace*{-.5cm} \sup_{\substack{\bphi_1, \bphi_2\\\|\mw\|,\|\blambda_0\|, \|\blambda_1\|\leq 1}}\hspace*{-.4cm} \bigg\{\int \trace (\bphi_1(x)-\kappa \blambda_0(x)) \hat d\bmu_1(x)\nonumber\\
&+\int \trace (\bphi_2(y)-\kappa \blambda_2(y))\hat d\bmu_2(y)\nonumber\\
&+\int \trace \left(d(x,y)\mw(x,y)-\bphi_1(x)-\bphi_2(y)\right)d\bm(x,y)\nonumber \\
&+\kappa \int \trace \bigg(\blambda_1(x) d\bmu_1(x)+\blambda_2(x) \bmu_2(x)\bigg)\bigg\}.\label{eq:infsupMatrixA}
\end{align}
\end{subequations}
Since this optimization problem is convex, the optimal value does not change by switching the $\inf$ and $\sup$ in \eqref{eq:infsupMatrixA}.
The optimal assignement for $\mw, \bphi_k, \blambda_k$ must satisfy
\begin{subequations}
\begin{eqnarray}
d(x,y)\mw(x,y)-\bphi_1(x)-\bphi_2(y)&=&0,\label{eq:a}\\
\bphi_1(x)-\kappa\blambda_0 (x)&=&0,\label{eq:b}\\
\bphi_2(y)-\kappa \blambda_1(y)&=&0.\label{eq:c}
\end{eqnarray}
\end{subequations}
Setting $x=y$ in \eqref{eq:a}, we obtain $\bphi_1=-\bphi_2=:\bphi$. Thus
\[
\bphi(x)-\bphi(y)\leq d(x,y)I, \text{~and~} \|\bphi\|\leq \kappa.
\]
By substituting these conditions to \eqref{eq:infsupMatrixA}, we derive \eqref{eq:dwk_dual_matrix}.

\bibliographystyle{IEEEtran}
\bibliography{IEEEabrv,TestFunction}

\begin{thebibliography}{10}
\providecommand{\url}[1]{#1}
\csname url@samestyle\endcsname
\providecommand{\newblock}{\relax}
\providecommand{\bibinfo}[2]{#2}
\providecommand{\BIBentrySTDinterwordspacing}{\spaceskip=0pt\relax}
\providecommand{\BIBentryALTinterwordstretchfactor}{4}
\providecommand{\BIBentryALTinterwordspacing}{\spaceskip=\fontdimen2\font plus
\BIBentryALTinterwordstretchfactor\fontdimen3\font minus
  \fontdimen4\font\relax}
\providecommand{\BIBforeignlanguage}[2]{{%
\expandafter\ifx\csname l@#1\endcsname\relax
\typeout{** WARNING: IEEEtran.bst: No hyphenation pattern has been}%
\typeout{** loaded for the language `#1'. Using the pattern for}%
\typeout{** the default language instead.}%
\else
\language=\csname l@#1\endcsname
\fi
#2}}
\providecommand{\BIBdecl}{\relax}
\BIBdecl

\bibitem{Ning2013matrix}
L.~Ning, T.~Georgiou, and A.~Tannenbaum, ``Matrix-valued {M}onge-{K}antorovich
  optimal mass transport,'' \emph{IEEE Transactions on Automatic Control}, to
  appear, 2015.

\bibitem{Ning2013}
L.~Ning, ``Matrix-valued optimal mass transportation and its applications,''
  Ph.D. dissertation, University of Minnesota, December 2013.

\bibitem{Zolotarev1984probability}
V.~Zolotarev, ``Probability metrics,'' \emph{Theory of Probability \& Its
  Applications}, vol.~28, no.~2, pp. 278--302, 1984.

\bibitem{Muller1997integral}
A.~M{\"u}ller, ``Integral probability metrics and their generating classes of
  functions,'' \emph{Advances in Applied Probability}, pp. 429--443, 1997.

\bibitem{rachev2013methods}
S.~T. Rachev, L.~Klebanov, S.~V. Stoyanov, and F.~Fabozzi, \emph{The methods of
  distances in the theory of probability and statistics}.\hskip 1em plus 0.5em
  minus 0.4em\relax Springer, 2013.

\bibitem{Rachev1991probability}
S.~T. Rachev, \emph{Probability metrics and the stability of stochastic
  models}.\hskip 1em plus 0.5em minus 0.4em\relax Wiley Chichester, 1991, vol.
  334.

\bibitem{Villani_book}
C.~Villani, \emph{Topics in optimal transportation}.\hskip 1em plus 0.5em minus
  0.4em\relax American Mathematical Society, 2003, vol.~58.

\bibitem{Jiang2012Geometric}
X.~Jiang, Z.~Luo, and T.~Georgiou, ``Geometric methods for spectral analysis,''
  \emph{IEEE Transactions on Signal Processing}, vol.~60, no.~3, pp.
  1064--1074, 2012.

\bibitem{Ning2013Covariance}
L.~Ning, X.~Jiang, and T.~Georgiou, ``On the geometry of covariance matrices,''
  \emph{IEEE Signal Processing Letters}, vol.~20, no.~8, pp. 787--790, 2013.

\bibitem{Georgiou2009metrics}
T.~Georgiou, J.~Karlsson, and M.~Takyar, ``Metrics for power spectra: an
  axiomatic approach,'' \emph{IEEE Transactions on Signal Processing}, vol.~57,
  no.~3, pp. 859--867, 2009.

\bibitem{Karlsson2012uncertainty}
J.~Karlsson and T.~Georgiou, ``Uncertainty bounds for spectral estimation,''
  \emph{IEEE Transactions on Automatic Control}, vol.~58, no.~7, pp.
  1659--1673, 2013.

\bibitem{Connes1995noncommutative}
A.~Connes, \emph{Noncommutative geometry}.\hskip 1em plus 0.5em minus
  0.4em\relax Academic press, 1995.

\bibitem{Rieffel1999metrics}
M.~Rieffel, ``Metrics on state spaces,'' \emph{Doc. Math., J. DMV}, vol.~4, pp.
  559--600, 1999.

\bibitem{DAndrea2010view}
F.~D'Andrea and P.~Martinetti, ``A view on optimal transport from
  noncommutative geometry,'' \emph{SIGMA}, vol.~6, no. 057, p.~24, 2010.

\bibitem{Ferrante_time}
A.~Ferrante, C.~Masiero, and M.~Pavon, ``{Time and spectral domain relative
  entropy: A new approach to multivariate spectral estimation},'' \emph{IEEE
  Transactions on Automatic Control}, vol.~57, no.~10, pp. 2561--2575, 2012.

\bibitem{Ferrante_hellinger}
A.~Ferrante, M.~Pavon, and F.~Ramponi, ``Hellinger versus {K}ullback--{L}eibler
  multivariable spectrum approximation,'' \emph{IEEE Transactions on Automatic
  Control}, vol.~53, no.~4, pp. 954--967, 2008.

\bibitem{Ferrante2012}
A.~Ferrante, M.~Pavon, and M.~Zorzi, ``A maximum entropy enhancement for a
  family of high-resolution spectral estimators,'' \emph{IEEE Transactions on
  Automatic Control}, vol.~57, no.~2, pp. 318--329, 2012.

\bibitem{Afsari2013}
B.~Afsari and R.~Vidal, ``The alignment distance on space of linear dynamical
  systems,'' in \emph{IEEE Conference on Decision and Control}, 2013.

\bibitem{JNG}
X.~Jiang, L.~Ning, and T.~Georgiou, ``Distances and {R}iemannian metrics for
  multivariate spectral densities,'' \emph{IEEE Transactions on Automatic
  Control}, vol.~57, no.~7, pp. 1723--1735, 2012.

\end{thebibliography}
\end{document}